\documentclass[10pt,final,journal,a4paper,oneside,twocolumn]{IEEEtran}

\usepackage{cite}
\usepackage{ulem}
\usepackage{soul,color}
\usepackage{srcltx}
\usepackage{multirow}
\usepackage[tbtags,sumlimits,nointlimits,reqno]{amsmath}
\allowdisplaybreaks[4]
\usepackage{graphicx}
\usepackage{stfloats}
\usepackage{subfigure}
\usepackage{psfrag}
\usepackage{color}
\usepackage{amssymb}

\begin{document}

\title{Single-Step Tunable Group Delay Phaser \\for Real-Time Spectrum Sniffing}

\author{%
       Tongfeng~Guo,
       Qingfeng~Zhang,
       Yifan Chen,
       Rui Wang,
       and~Christophe~Caloz
\thanks{Manuscript received Jan. 23, 2015.}
\thanks{T.~Guo, Q.~Zhang, Y. Chen and R. Wang are with the Department of Electrical and Electronic Engineering, South University of Science and Technology of China, Shenzhen, Guangdong, China, Email: zhang.qf@sustc.edu.cn.

C. Caloz is with the Department of Electrical
Engineering, Poly-Grames Research Center, \'{E}cole Polytechnique de Montr\'{e}al,
Montr\'{e}al, QC, Canada H3T 1J4. C. Caloz is also with King Abdulaziz University, Jeddah, Saudi Arabia.
}
}

\maketitle


\begin{abstract}

This paper presents a single-step tunable group delay phaser for spectrum sniffing. This device may be seen as a ``time filter'', where frequencies are suppressed by time separation rather than by spectral attenuation. Compared to its multiple-step counterpart, this phaser features higher processing resolution, greater simplicity, lower loss and better channel equalization, due to the smaller and channel-independent group delay swing. A three-channel example is provided for illustration.

\end{abstract}


\IEEEpeerreviewmaketitle

\section{Introduction}
Real-time analog signal processing (R-ASP), inspired by ultrafast optics and surface acoustics, exhibits remarkable advantages over digital signal processing in some applications~\cite{caloz2013mm}, such as for instance fast and low-cost spectrum analysis~\cite{Laso_TMTT_03_2003,Schwartz_MWCL_04_2006,gupta2009microwave}, pulse compansion~\cite{Azana_TMTT_2007s} and spectrum sniffing~\cite{Nikfal}.

Real-time spectrum sniffer, as a typical R-ASP system, is particularly promising for cognitive radio~\cite{cognitive_radio}. The key component of the real-time spectrum sniffer is a phaser, which exhibits a stair-case group delay response versus frequency~\cite{Nikfal}. Different temporal frequencies of the multi-tone input pulse are subsequently separated in time due to the different delays they experience. However, the design complexity of the stair-case phaser greatly increases as the channel number increases.

We propose here a single-step tunable group delay phaser for enhanced real-time spectrum sniffing. This phaser features lower design complexity and improved system performance compared to~\cite{Nikfal}.
\section{Principle}\label{sec:principle}
\begin{figure}[!t]
\centering
  \psfrag{x}[c][c]{\footnotesize $\omega$}
  \psfrag{y}[c][c]{\footnotesize $\tau$}
  \psfrag{1}[c][c]{\footnotesize $\tau_{1}$}
  \psfrag{2}[c][c]{\footnotesize $\tau_{2}$}
  \psfrag{3}[c][c]{\footnotesize $\tau_{i}$}
  \psfrag{4}[c][c]{\footnotesize $\tau_{N}$}
  \psfrag{5}[c][c]{\footnotesize $\tau_{5}$}
  \psfrag{6}[c][c]{\footnotesize $\tau_{6}$}
  \psfrag{7}[c][c]{\footnotesize $\tau_{1}$}
  \psfrag{a}[c][c]{\footnotesize $\omega_{1}$}
  \psfrag{b}[c][c]{\footnotesize $\omega_{2}$}
  \psfrag{c}[c][c]{\footnotesize $\omega_{n}$}
  \psfrag{d}[c][c]{\footnotesize $\omega_{N}$}
  \psfrag{m}[c][c]{\footnotesize $(a)$}
  \psfrag{n}[c][c]{\footnotesize $(b)$}
  \psfrag{g}[c][c]{\footnotesize reconfigurable state}
  \includegraphics[width=6.5cm]{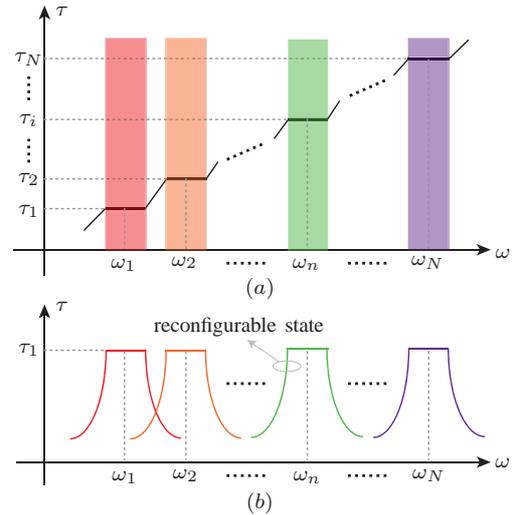}\\
  \caption{Comparison of the group delay responses of (a)~a stair-case group delay phaser~\cite{Nikfal} and (b)~the proposed \emph{single-step tunable} group delay phaser for an $N-$channel spectrum sniffer.}\label{fig:fig11}
\end{figure}

Figure~\ref{fig:fig11}(a) shows an $N$-channel stair-case group delay response, where different delays $\tau_1,\tau_2,\ldots,\tau_N$ corresponds to different channels centered at $\omega_{1},\omega_{2},\ldots,\omega_{N}$. The signals in different channels experience different delays through the phaser, and therefore get separated in time. For complete separation, the delay step should comply with the uncertainty principle~\cite{uncertainty}: $\tau_n-\tau_{n-1} \geq 2\pi/[\omega_{n}-\omega_{(n-1)}]$. The main benefit of stair-case group delay sniffing is its instantaneous channel discrimination following from the fact that all the channels are processed simultaneously. However, this approach involves challenging phaser design due to increasing group-delay-bandwidth product with increasing number of channels. Increasing group delay also leads to increasing loss, as loss is inversely proportional to group delay, and therefore harder channel discrimination due to lower signal-to-noise ratio.

The proposed single-step tunable group delay response, shown in Fig.~\ref{fig:fig11}(b), overcomes the aforementioned issues. It is composed of a number of reconfigurable states corresponding to the number of channels, each of which exhibits a single step group delay response centered at frequency $\omega_n$ and therefore discriminating channel~$n$ from the others. In contrast to the staircase case, the average group delay is the same for all the channels, so that all the channels are on an equal footing from the viewpoint of loss, if one (reasonably) ignores the loss difference due to frequency-varying dielectric and ohmic losses. The single-step tunable response also leads to simpler phaser design for a given resolution since each single step in Fig.~\ref{fig:fig11}(a) can easily be designed to provide a larger group delay swing than the individual steps in Fig.~\ref{fig:fig11}(a). The only disadvantage of the proposed approach lies in the limited speed due to the requirement of state switching; it is thus only a \emph{quasi real-time} technique.

\begin{figure}[!t]
\centering
  \centering
  \psfrag{y}[c][c]{\footnotesize $|S_{21}|$}
  \psfrag{w}[c][c]{\footnotesize $\omega_{2}$}
  \psfrag{l}[c][c]{\footnotesize $\omega_{1}$}
  \psfrag{m}[c][c]{\footnotesize $\omega_{3}$}
  \psfrag{x}[c][c]{\footnotesize $\omega$}
  \psfrag{d}[c][c]{\footnotesize $\omega_{2}$}
  \psfrag{e}[c][c]{\footnotesize $\omega_{1},\omega_{3}$}
  \psfrag{a}[c][c]{\footnotesize $\omega_{1},\omega_{2},\omega_{3}$}
  \psfrag{b}[c][c]{\footnotesize \textcolor{red}{Device}}
  \psfrag{c}[c][c]{\footnotesize $t$}
  \psfrag{z}[c][c]{\footnotesize $\tau_{21}$}
  \psfrag{1}[c][c]{\footnotesize \textcolor[RGB]{126,128,51}{Input Signal}}
  \psfrag{2}[c][c]{\footnotesize Filtering Device}
  \psfrag{3}[c][c]{\footnotesize \textcolor[RGB]{15,135,84}{Output Signal}}
  \psfrag{4}[c][c]{\footnotesize Filter}
  \psfrag{6}[c][c]{\footnotesize flat delay}
  \psfrag{5}[c][c]{\footnotesize Phaser}
  \includegraphics[width=8cm]{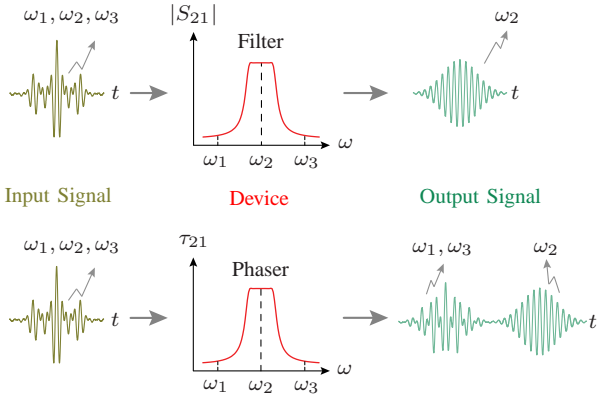}\\
  \caption{Comparison between the operation principles of a filter, essentially featuring a selective magnitude response, and the proposed phaser, essentially featuring a selective delay response.}\label{fig:fig2}
\end{figure}

The single-step group delay response of a given state in Fig.~\ref{fig:fig11}(b) is very similar to the magnitude response of a magnitude filter. However, the two devices operate in a totally different fashion, as illustrated by Fig.~\ref{fig:fig2}. The figure uses a three-tone pulse and bandpass filter example, where pulses of central frequencies $\omega_{1},\omega_{2},\omega_{3}$ are presented to the bandpass filter; energies centered at $\omega_{1}$ and $\omega_{3}$ are attenuated whereas that of $\omega_{2}$ is transmitted, as shown in the top part of the figure. In contrast, the phaser delays the unwanted spectral content by a lower amount than the wanted contents instead of suppressing it, as shown in the bottom part. The spectrum centered at $\omega_{2}$ is delayed more than the spectra around $\omega_{1}$, leading to time discrimination. The undesired channels may then be suppressed using dynamic thresholding or time gating.
\section{Phaser Design}\label{sec:methodology}
Several types of phasers, including reflection type~\cite{Zhang2012}, cross-coupled transmission type~\cite{Zhang_TMTT_cross} and all-pass C-section type ~\cite{Gupta-allpass}, can provide the group delay responses in Fig.~\ref{fig:fig11}. Here, we use a reflection-type phaser implemented in microstrip technology, featuring simple architecture and easy tuning compared with cross-coupled and C-section phasers.
\subsection{Static Case}
\begin{figure}[!t]
  \centering
  \psfrag{a}[c][c]{\footnotesize $J_{1}$}
  \psfrag{b}[c][c]{\footnotesize $J_{2}$}
  \psfrag{c}[c][c]{\footnotesize $J_{3}$}
  \psfrag{d}[c][c]{\footnotesize $C_{1}$}
  \psfrag{e}[c][c]{\footnotesize $C_{2}$}
  \psfrag{f}[c][c]{\footnotesize $C_{n}$}
  \psfrag{g}[c][c]{\footnotesize $\lambda$/4}
  \psfrag{h}[c][c]{\footnotesize $\lambda$/2}
  \psfrag{i}[c][c]{\footnotesize $J_{n}$}
  \psfrag{x}[c][c]{\footnotesize $(a)$}
  \psfrag{y}[c][c]{\footnotesize $(b)$}
  \psfrag{m}[l][c]{\footnotesize O.C.}
  \includegraphics[width=8cm]{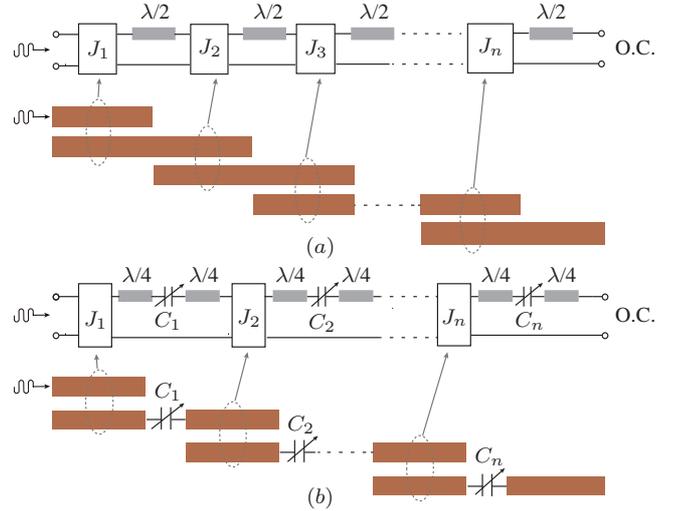}\\
  \caption{Circuit model and microstrip reflection-type implementation of the proposed phaser: (a)~static case, (b)~tunable case.}\label{fig:fig3}
\end{figure}

Figure~\ref{fig:fig3}(a) shows the J-inverter circuit model of a reflection-type phaser, which is a dual circuit of that used in~\cite{Zhang2012}. The J-inverter and the open end in Fig.~\ref{fig:fig3}(a) correspond to the K-inverter and short end in~\cite{Zhang2012}, respectively. All the inverters are connected by half-wavelength transmission lines, which act as resonators. The circuit is implemented by edge-coupled transmission lines, where edge coupling is modeled by the J-inverter~\cite{pozar2009microwave}. The closed-form synthesis method of~\cite{Zhang2012} can be directly applied to the circuit in Fig.~\ref{fig:fig3}(a). However, the circuit in Fig.~\ref{fig:fig3}(a) is a static circuit, and should therefore be slightly modified to achieve tunability.
\subsection{Tunable Case}
\begin{figure}[!t]
  \centering
  \psfrag{a}[c][c]{\footnotesize $\omega_{0}$}
  \psfrag{b}[c][c]{\footnotesize $\omega_{1}$}
  \psfrag{c}[c][c]{\footnotesize $\omega_{2}$}
  \psfrag{d}[c][c]{\footnotesize $2\omega C_{1} Z_{0}$}
  \psfrag{e}[c][c]{\footnotesize $2\omega C_{2} Z_{0}$}
  \psfrag{f}[c][c]{\footnotesize $-\tan\left(\dfrac{\pi\omega}{\omega_0}\right)$}
  \psfrag{g}[c][c]{\footnotesize $\omega_{2}<\omega_{1}<\omega_{0}$}
  \psfrag{h}[c][c]{\footnotesize $C_{2}>C_{1}>0$}
  \psfrag{i}[c][c]{\footnotesize $J$}
  \psfrag{j}[c][c]{\footnotesize $\lambda/4$}
  \psfrag{k}[c][c]{\footnotesize $C$}
  \psfrag{o}[l][c]{\footnotesize functions}
  \psfrag{0}[l][c]{\footnotesize $0$}
  \psfrag{x}[c][c]{\footnotesize $\omega$}
  \psfrag{p}[c][c]{\footnotesize $Y_\text{in}$}
  \psfrag{q}[c][c]{\footnotesize $\frac{\omega_0}{2}$}
  \includegraphics[width=8.6cm]{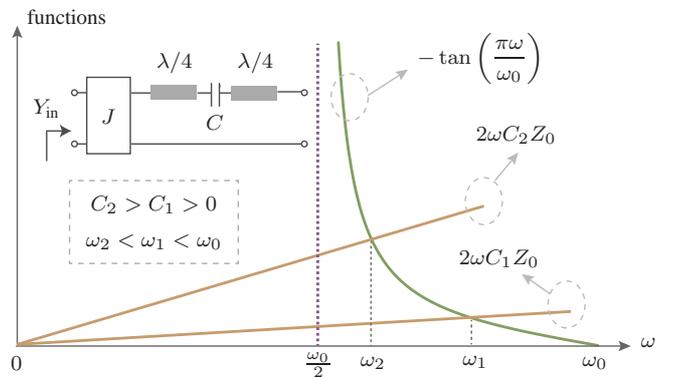}\\
  \caption{Effect of capacitance variation on the resonant frequency of the loaded transmission line resonators in Fig.~\ref{fig:fig3}.}\label{fig:even_odd}
\end{figure}

Figure~\ref{fig:fig3}(b) shows the circuit model and implementation of the tunable version of the phaser, where varactors have been added to the middle part of the half-wavelength transmission line. These varactors, biased by different voltages, provide tunable capacitances, which results in tunable resonant frequencies. The relation between the inserted capacitance and the resonant frequency, $\omega_0$, is computed by analyzing the input admittance $Y_\text{in}$ of the resonator element in the inset figure of Fig.~\ref{fig:even_odd}. The resonance occurs when $Y_\text{in}=0$, which leads to
  \begin{align}
-\tan\left(\dfrac{\pi\omega}{\omega_{0}}\right)=2\omega C Z_{0}.~\label{eq:eq1}
\end{align}
\noindent No closed-form solution is available for the above equation. However, it can be solved graphically by finding the interception point of two curves corresponding to the left-hand and right-hand sides of~\eqref{eq:eq1}, as shown in Fig.~\ref{fig:even_odd}. Note that as the capacitance increases, the resonance frequency decreases. The resonant frequency is thus inversely proportional to the capacitance. Also note that the resonant frequency is nonlinearly related to the capacitance, due to the nonlinear functions constituting the left-hand side of~\eqref{eq:eq1}. This nonlinear effect is increasingly notable for increasing capacitance.

The design procedure comprises the following steps. Firstly, one synthesizes the static circuit in Fig.~\ref{fig:fig3}(a) for a specified single-step group delay in one channel using the closed-form technique provided in~\cite{Zhang2012}. One subsequently applies the synthesized parameters to Fig.~\ref{fig:fig3}(b) with a set of chosen capacitances. One may locally tune and optimize the initial parameters and the capacitances to better fit the specified group delay response. The same procedure is repeated for all the channels with the same circuital parameters but different sets of capacitances.
\section{Design Example}\label{sec:example}
To illustrate the proposed system, consider a three-channel example, with $20$~MHz channel bandwidths centered at $2.3$, $2.5$, and $2.7$~GHz. For these parameters, the group delay step should be larger than $5$~ns, according to the uncertainty principle~\cite{uncertainty}. For proof of concept, we employ three groups of static capacitors instead of varactors.

\begin{figure}[!t]
  \centering
  \psfrag{a}[c][c]{\footnotesize Unit: mm}
  \includegraphics[width=8cm]{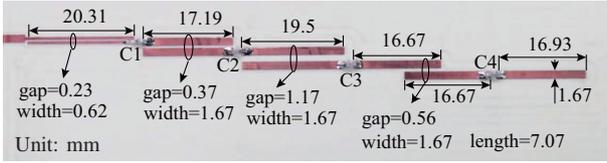}\\
  \caption{Photograph (top view) of the fabricated prototype.}\label{fig:fig4}
\end{figure}

\begin{figure}[!t]
  \centering
  \psfrag{a}[l][c]{\footnotesize Circuit}
  \psfrag{b}[l][c]{\footnotesize Full-wave}
  \psfrag{c}[l][c]{\footnotesize Measured}
  \psfrag{d}[c][c]{\footnotesize State 1}
  \psfrag{e}[c][c]{\footnotesize State 2}
  \psfrag{f}[c][c]{\footnotesize State 3}
    \psfrag{l}[c][c]{\footnotesize \shortstack{$f_{1}$\\channel}}
  \psfrag{m}[c][c]{\footnotesize \shortstack{$f_{2}$\\channel}}
  \psfrag{n}[c][c]{\footnotesize \shortstack{$f_{3}$\\channel}}
  \psfrag{x}[c][c]{\footnotesize Frequency (GHz)}
  \psfrag{y}[c][c]{\footnotesize Group delay (ns)}
  \includegraphics[width=6.5cm]{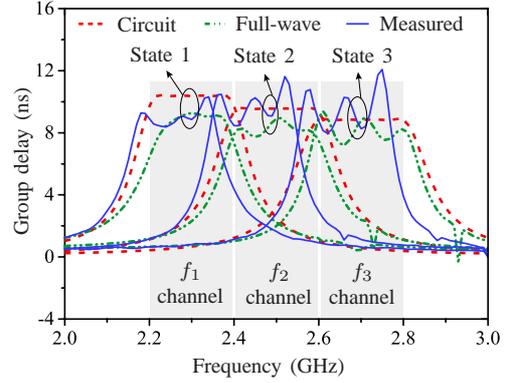}\\
  \caption{Full-wave and measured group delay responses of the proposed phaser ($f_{1}=2.3~\text{GHz}, f_{2}=2.5~\text{GHz}, f_{3}=2.7~\text{GHz}$).}\label{fig:fig5}
\end{figure}

\begin{figure}[!t]
  \centering
  \psfrag{a}[c][c]{\footnotesize $V_\text{in}$}
  \psfrag{b}[c][c]{\footnotesize Phaser in Fig.~\ref{fig:fig5}}
  \psfrag{c}[c][c]{\footnotesize $V_\text{out}$}
  \psfrag{d}[c][c]{\footnotesize Temporal Filter}
  \psfrag{e}[c][c]{\footnotesize $f'_{1}$}
  \psfrag{f}[c][c]{\footnotesize $f'_{2}$}
  \psfrag{g}[c][c]{\footnotesize $f'_{3}$}
  \psfrag{h}[c][c]{\footnotesize $f'_{1},f'_{2},f'_{3}$}
  \psfrag{i}[c][c]{\footnotesize $f'_{1}:2.25~\text{GHz}$}
  \psfrag{j}[c][c]{\footnotesize $f'_{2}:2.45~\text{GHz}$}
  \psfrag{k}[c][c]{\footnotesize $f'_{3}:2.65~\text{GHz}$}
  \psfrag{l}[c][c]{\footnotesize state 1}
  \psfrag{m}[c][c]{\footnotesize state 2}
  \psfrag{n}[c][c]{\footnotesize state 3}
  \psfrag{x}[c][c]{\footnotesize Time (ns)}
  \psfrag{y}[c][c]{\footnotesize $V_\text{out}$ (v)}
  \psfrag{z}[c][c]{\footnotesize $V_\text{in}$ (v)}
  \includegraphics[width=8.6cm]{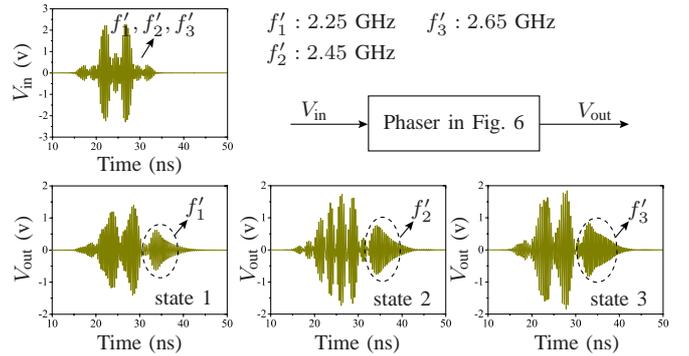}\\
  \caption{Responses of the spectrum sniffer based on the fabricated phaser shown in Fig.~\ref{fig:fig4} and characterized in Fig.~\ref{fig:fig5}. A channel frequencies shift of $0.05~\text{GHz}$ was introduced for alignment and comparison with experiments.}\label{fig:fig6}
\end{figure}

Figure~\ref{fig:fig4} shows the fabricated prototype, where the substrate is Rogers RO4350 ($\varepsilon_{r}=3.66$,$\tan\delta=0.004$) with thickness $0.762$~mm, and the capacitors are GQM18 series from Murata Corporation. The synthesized capacitor values are listed in Tab.~\ref{table1}. The synthesized and measured group delay responses are shown in Fig.~\ref{fig:fig5}. Note that the measured results are in reasonable agreement with full-wave and circuit results, despite a slight frequency shift ($\approx 0.05~\text{GHz}$) due to fabrication tolerance and capacitor parasitic effects. Also note the related ripples in the measured group delays. Fortunately, as shown in Fig.~\ref{fig:fig6} which uses the measured group delay responses, the effect of these ripples on the overall system is of no significant in spectrum sniffing: the desired channels are perfectly discriminated, and therefore detected, despite slightly distorted pulse shapes.
\begin{table}[!t]
\renewcommand{\arraystretch}{1}
\caption{Capacitances in different states (unit:pF).}
\label{table1}
\centering
\begin{tabular*}{0.3\textwidth}{c|c|c|c|l}
  \hline
  \hline
  Operation State & $C_{1}$ & $C_{2}$ & $C_{3}$ & $C_{4}$ \\
  \hline
  state 1 &51&47&75&10 \\
  \hline
  state 2 &4&5&5&6.2 \\
  \hline
  state 3 &2&2&2&2.2 \\
  \hline
  \hline
\end{tabular*}
\end{table}
\section{Conclusion}\label{sec:conclusion}
A single-step tunable group delay phaser has been proposed and demonstrated experimentally in a three-channel spectrum sniffer.
\bibliographystyle{IEEEtran}
\bibliography{Guobib}
\end{document}